\documentclass[prl,amsmath, twocolumn, superscriptaddress,showpacs]{revtex4-1}
\usepackage{graphicx}
\usepackage{color}

\begin{document}

\title{Imitating Chemical Motors with Optimal Information Motors}

\author{Jordan M.\ Horowitz}
\affiliation{Departamento de F\'{\i}sica At\'omica, Molecular y Nuclear and GISC, Universidad Complutense de Madrid, 28040 Madrid, Spain}

\author{Takahiro Sagawa}
\affiliation{The Hakubi Center for Advanced Research, Kyoto University, Yoshida-ushinomiya cho, Sakyo-ku, Kyoto 606-8302, Japan}
\affiliation{Yukawa Institute for Theoretical Physics, Kyoto University, Kitashirakawa-oiwake cho, Sakyo-ku, Kyoto 606-8502, Japan}

\author{Juan M.\ R.\ Parrondo}
\affiliation{Departamento de F\'{\i}sica At\'omica, Molecular y Nuclear and GISC, Universidad Complutense de Madrid, 28040 Madrid, Spain}

\date{\today}

\begin{abstract}
To induce transport, detailed balance must be broken.
A common mechanism is to bias the dynamics with a thermodynamic fuel, such as chemical energy.
An intriguing, alternative strategy is for a Maxwell demon to effect the bias using feedback.
We demonstrate that these two different mechanisms lead to  distinct thermodynamics by contrasting a chemical motor and information motor with identical dynamics.
To clarify this difference,  we study both models within one unified framework, highlighting the role of the interaction between the demon and the  motor. 
This analysis elucidates the manner in which  information is incorporated into a physical system.
\end{abstract}


\maketitle

Information is physical~\cite{Landauer1991}: it is stored in physical memories, and therefore the processing of information is constrained by the same thermodynamic limitations as any other physical process~\cite{Leff, Maruyama2009}.
Remarkably, once  information has been obtained, it can then serve as a thermodynamic resource similar to free energy. 
These observations have been the  historic points of departure for investigations into the nature of information~\cite{Leff, Maruyama2009}.
As a consequence, research has either focused on the manipulation of information in isolated memories, or simply on the engines that utilize that information. 
This division has been fruitful.
Theoretical studies of memories, which have been verified by experiment~\cite{Berut2011}, have led to insights into the thermodynamic costs of measurement~\cite{Granger2011} and erasure~\cite{Bennett:1982wx,Piechocinska2000,Landauer:2006gs,Dillenschneider2009,delRio2010,Sagawa2009}; copying~\cite{Bennett:1979tb,Andrieux:2008tp}; and proofreading~\cite{Bennett:1979tb}; while theoretical \cite{Cao2004,Suzuki2009,Vaikuntanathan2011,Abreu2011,Bauer2012,Horowitz2011,Horowitz:2011vg} and experimental~\cite{Lopez2008,Toyabe2010} investigations of information (or feedback) motors have  explored the fundamental  limits to the conversion of information into work.

Nevertheless, the thermodynamic qualities of information still need to be clarified, especially the mechanisms that allow a motor to exploit  information  to rectify thermal fluctuations.
With this goal in mind,  we highlight in this Letter the difference between information and a more traditional thermodynamic resource, the chemical free energy.
We elaborate this distinction by comparing the entropy production rates of two motors with  \emph{identical} dynamics: a chemical motor powered by a chemical potential  gradient and an information motor driven by feedback.
For the chemical motor, we use traditional methods of thermodynamic analysis~\cite{Parrondo2002}.
However, such methods cannot be applied to the information motor when the memory is left unspecified.
Even still, a useful bound for its entropy production can be obtained from a refinement of the second law of thermodynamics for feedback and information, introduced by Sagawa and Ueda~\cite{Sagawa2009}.
We demonstrate that, despite the identical dynamics, the information motor presents qualitatively different thermodynamics.
We then trace this discrepancy to the features of the interaction between the ratchet and the memory by introducing a physical model of the information motor.

\begin{figure}[htb]
\[
\includegraphics[trim=3cm .4cm 3cm .4cm, clip=true, scale=.24]{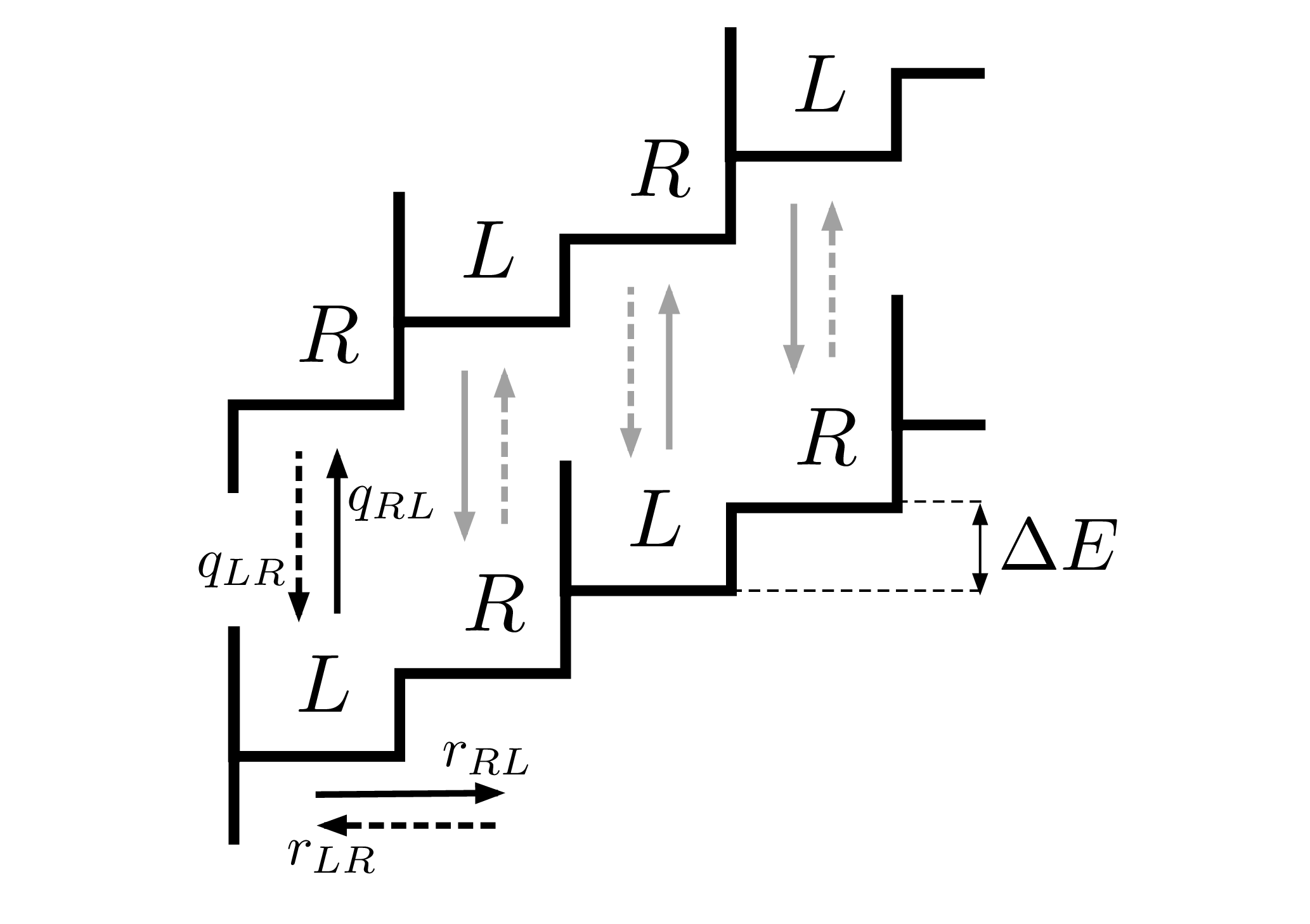}\ \qquad
\includegraphics[trim=.4cm 4cm .8cm 4.6cm, clip=true, scale=.16]{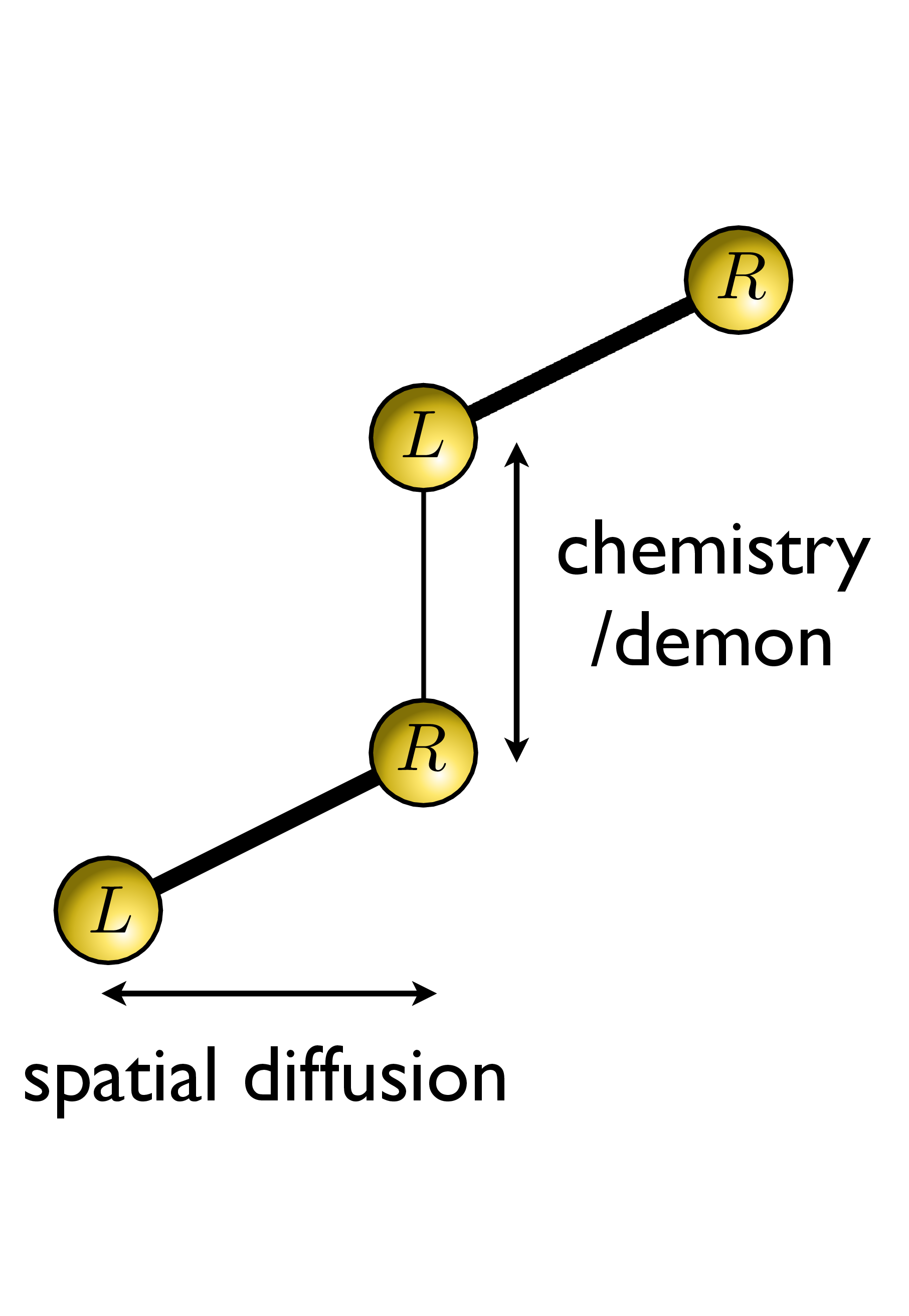}
\]
\caption{
Depiction of the Brownian ratchet template.
A particle moves in a periodic potential of period $l$ against a force $F=2\Delta E/l$.
The potential switches randomly between two configurations mediated by either a chemical reaction (chemical motor) or a demon (information motor).
Spatial diffusion and potential switches induce $R\leftrightarrow L$ transitions  at rates $r_{ij}$ and $q_{ij}$ ($i,j=R,L$), respectively.
}\label{figmotor}
\end{figure}

Our motors are patterned on the Brownian ratchet~\cite{Reimann2002} pictured in Fig.~\ref{figmotor}. 
The ratchet is composed of a particle driven against a force $F$  by  a flashing periodic potential.
The potential fluctuates between two configurations consisting of a series of offset infinite barriers that confine the particle to boxes of length $l$.
Within each box the particle has two spatial states: $z=L$ (left) and a higher energy state $R$ (right) with energy difference $\Delta E=Fl/2$,  where energy units are set by fixing the temperature, $kT=1$.
The probability $p_z(t)$ to be in state $z=R,L$ at time $t$ obeys the master equation~\cite{VanKampen}
\begin{equation}\label{eq:master}
\begin{split}
\dot{p}_R(t)&=(r_{RL}+q_{RL})p_L(t)-(r_{LR}+q_{LR})p_R(t) \\ 
\dot{p}_L(t)&=(r_{LR}+q_{LR})p_R(t)-(r_{RL}+q_{RL})p_L(t).
\end{split}
\end{equation}
Here, diffusive jumps \emph{within} each box are thermally activated with rates $r_{LR}$ ($r_{RL}$) from  $R$ to $L$ ($L$ to $ R$) that verify detailed balance, $r_{LR}/r_{RL}=e^{\Delta E}$~\cite{Parrondo2002}; whereas transitions effected by potential switches occur with rates $q_{LR}$ ($q_{RL}$) for  $R\to L$ ($L\to R$) and are induced by a \emph{different mechanism in each motor}, see Fig.~\ref{figmotor}.
Furthermore, these switches do not require  energy, as we assume that $R$ ($L$) in the lower configuration in Fig.~\ref{figmotor} has the same energy as $L$ ($R$) in the upper.
The ratchet functions as long as in the stationary state ($\dot{p}_R(t)=\dot{p}_L(t)=0$) the $R\to L$ switches  occur more often than the reverse, $L \to R$.
In which case, the stationary current $J $ -- the average net number of jumps per unit time against the load -- is positive,  and work is extracted at a rate~\cite{Parrondo2002,SuppMat}
\begin{equation}\label{eq:workExt}
{\dot W}_{\rm ext}=J\Delta E.
\end{equation}

It will prove convenient in our subsequent calculations to assume that potential switches are slow compared to the spatial transitions ($q\ll r$).
In this limit, the stationary  solution of Eq.~(\ref{eq:master})~\cite{SuppMat}, 
\begin{equation}
\label{pr}
p_R=\frac{1}{1+e^{\Delta E}},\qquad p_L=\frac{e^{\Delta E}}{1+e^{\Delta E}},
\end{equation}
is in equilibrium with respect to diffusion, inducing a current $J=q_{LR}p_R-q_{RL}p_L$.

In the chemical motor, the potential switches are biased by coupling them to an out-of-equilibrium chemical reaction between species $A$ and $B$ through the formula $R+A\leftrightarrow L+B$.
Detailed balance enforces that the chemical potential  difference between $A$ and $B$, $\Delta \mu\equiv\mu_A-\mu_B >0$, satisfies $\Delta \mu=\ln (q_{LR}/q_{RL})$.
The resulting scheme corresponds to the minimal tight-coupled chemical motor extensively used to model protein motors~\cite{Parrondo2002}.
When $\Delta\mu>\Delta E$, ${\dot W}_{\rm ext}$ [Eq.~(\ref{eq:workExt})] is extracted by consuming chemical free energy per unit time $\dot F_{\rm chem}= J\Delta\mu$.
The resulting entropy production rate is~\cite{Parrondo2002}
\begin{equation}\label{schem}
\dot S^{\rm (chem. mot.)}={\dot F}_{\rm chem} - {\dot W}_{\rm ext}=J(\Delta \mu-\Delta E)\ge 0.
\end{equation}
 
To contrast with the chemical motor, we now consider an  information motor driven by feedback implemented by a device, or so-called demon, that switches the potential in response to measurements of the particle's position.

In order for the information motor to reproduce the stochastic potential switches of the ratchet, the demon  measures at random times according to a Poisson process with rate $\alpha=q_{RL}+q_{LR}$~\cite{VanKampen}.
This  scheme may be interpreted as the demon attempting to make a measurement in each small interval of time $\delta t$, but only succeeding with probability $\alpha \delta t\ll 1$.
When the demon succeeds, it measures $R$ or $L$ with a symmetric error, mistaking $R$ ($L$) for $L$ ($R$) with probability  $\epsilon=q_{RL}/(q_{RL}+q_{LR})$, and flips the potential when the outcome is $R$.
Moreover, the demon records the sequence of potential switches in a memory with states $m$, though for now we leave unspecified  the recording mechanism.
When the demon fails to make a measurement or the outcome is $L$, the memory is put in state $m=N$ for no-switch; whereas, when the outcome is $R$, the memory is set to $S$ for switch.

With this setup, potential switches occur at rates $q_{LR}=\alpha (1-\epsilon)$ and $q_{RL}=\alpha\epsilon$, as desired.
Comparison with the chemical motor leads to the correspondence  $\Delta \mu=\ln[(1-\epsilon)/\epsilon]$: $\epsilon=0$ is equivalent to $\Delta\mu=\infty$, and $\epsilon=1/2$ corresponds to an equilibrium fuel, $\Delta\mu=0$.

Even though the physical nature of the demon is unspecified, we can still discuss the information motor's thermodynamics  using the framework developed in Refs.~\cite{Sagawa2008,Sagawa2010,Ponmurugan2010,Abreu2012,Horowitz2010,Suzuki2010,Sagawa2009,Horowitz2011,Sagawa2012}.
Later, we  validate this approach by providing an explicit physical model for the demon.
The framework's main features can be simply obtained by introducing a nonequilibrium free energy~\cite{Esposito2011,Deffner:2012tm} for a system  whose mesoscopic states $x$ are in local equilibrium~\cite{Hill}.
To each system configuration $X=\{p_x,F_x\}$, characterized by free energy $F_x$ of state  $x$ and  probability $p_x$ to be in $x$, we assign a \emph{nonequilibrium free energy}   ($kT=1$)
\begin{equation}\label{eq:free}
{\cal F}(X)= \sum_x p_x F_x-H(X)\equiv F(X)-H(X), 
\end{equation}
 where $H(X)$ is the Shannon entropy~\cite{Cover}.
We call $F(X)=\sum p_x F_x$ the {\em bare free energy}. 
In equilibrium, $p^{\rm eq}_x=e^{-F_x}/Z$ with $Z=\sum e^{-F_x}$, and we recover ${\mathcal F}^{\rm eq}=-\ln Z$. 
The utility of ${\mathcal F}$ stems from the observation that the (irreversible) entropy production in a transition between  configurations is the amount by which the work $W$ exceeds the increment in the nonequilibrium free energy $\Delta {\mathcal F}$~\cite{Esposito2011,Deffner:2012tm}: 
\begin{equation}\label{eq:SinfoF}
\Delta_{i} {\mathcal S}=W-\Delta {\mathcal F}\ge 0.
\end{equation}
 While $\Delta_i{\mathcal S}$ only equals the change in thermodynamic  entropy for transitions between equilibrium states, away from equilibrium Eq.~(\ref{eq:SinfoF}) still offers useful  insight.
It bounds the work required for any process and can be shown to be a measure of irreversibility. Moreover, when a process connecting equilibrium states can be divided into different stages connecting nonequilibrium configurations, the sum of $\Delta_i{\mathcal S}$ over all these stages yields the total change in  equilibrium entropy.

We are interested in the entropy production for the coupled memory and ratchet, $X=(M,Z)$,  during measurement and feedback.
An ideal classical measurement correlates the initially uncorrelated memory and ratchet, ${\mathcal F}(M,Z)={\mathcal F}(M)+{\mathcal F}(Z)$, through an isothermal process, $(M,Z)\to(M^\prime, Z^\prime)$, \emph{without} affecting the ratchet, $Z^\prime=Z$, though possibly changing the nonequilibrium free energy of the  memory to ${\mathcal F}(M^\prime)\neq{\mathcal F}(M)$~\cite{Sagawa2009,Granger2011}. If before and after the measurement the bare free energies are additive, $F(M,Z)=F(M)+F(Z)$, then the nonequilibrium free energy can be cast into
\begin{equation}\label{eq:freeJoint}
{\cal F}(M^\prime, Z^\prime)={\cal F}(M^\prime)+{\cal F}(Z)+I(M^\prime,Z), 
\end{equation}
where the mutual information
\begin{equation}\label{eq:I}
I(M^\prime,Z)\equiv H(M^\prime)+H(Z)-H(M^\prime,Z)
\end{equation} 
measures correlations, satisfying $I\ge 0$ with $I=0$ only when $M^\prime$ and $Z$ are independent~\cite{Cover}.
Consequently, the creation of correlations, or measuring, increases the nonequilibrium free energy, requiring work $W_{\rm meas}$ and producing entropy according to Eqs.~(\ref{eq:SinfoF}) and (\ref{eq:freeJoint})~\cite{Sagawa2009,Granger2011,Sagawa2012},
\begin{equation}\label{eq:entMeas}
\begin{split}
\Delta_i {\mathcal S}_{\rm meas} &=W_{\rm meas}-\Delta {\mathcal F}(M,Z) \\
& = W_{\rm meas}-\Delta {\mathcal F}(M)-I(M^\prime,Z)\ge 0,
\end{split}
\end{equation}
where $\Delta {\mathcal F}(Y)={\mathcal F}(Y^\prime)-{\mathcal F}(Y)$.

Once the correlations have been established, they can be exploited through a subsequent isothermal process, $(M^\prime,Z^\prime)\to(M^{\prime\prime}, Z^{\prime\prime})$, that extracts work $W_{\rm ext}$ from the ratchet without altering the memory, ${\mathcal F}(M^{\prime\prime})={\mathcal F}(M^\prime)$. 
When all the correlations are removed [$I(M^{\prime\prime},Z^{\prime\prime})=0$], we  call this scenario feedback.
For the cyclic processes we consider here, ${\mathcal F}(Z^{\prime\prime})={\mathcal F}(Z)$, and the entropy production is [Eq.~(\ref{eq:SinfoF})]~\cite{Sagawa2008,Horowitz2010,Esposito2011,Sagawa2012}
\begin{equation}\label{eq:2LawFeed}
\Delta_i {\mathcal S}_{\rm fb}=I(M^\prime,Z)- W_{\rm ext}\ge 0.
\end{equation}
Only when the measurement is reversible ($\Delta_i {\mathcal S}_{\rm meas}=0$) does Eq.~(\ref{eq:2LawFeed}) represent the total  entropy production for the entire measurement and feedback cycle.
In general, $\Delta_i {\mathcal S}_{\rm fb}$ is only a lower bound.

Now, since the information motor utilizes feedback, we can use Eq.~(\ref{eq:2LawFeed}) to calculate its (minimum) entropy production rate. 
To this end, we  calculate the mutual information. 
The fast diffusion implies that the ratchet begins each $\delta t$ with the same equilibrium probability density~[Eq.~(\ref{pr})], independent of past measurements.
Consequently, each interval is independent and can be analyzed separately.
We  then obtain $I$ by substituting the probability density $p^\prime_{z,m}$ for the composite system after the measurement  into Eq.~(\ref{eq:I})~\cite{SuppMat} 
\begin{equation}\label{eq:infoRate}
{\dot I}=\frac{I(M^\prime,Z)}{\delta t} \simeq  p_Rq_{LR}\ln\frac{q_{LR}}{q_S}+p_Lq_{RL}\ln \frac{q_{RL}}{q_S},
\end{equation}
where $q_S=(p_{R,S}^\prime+p_{L,S}^\prime)/\delta t$ is the switching rate.
Then by combining Eqs.~(\ref{eq:workExt}), (\ref{eq:2LawFeed}), and (\ref{eq:infoRate}), we find the  entropy production rate
\begin{equation}\label{sinfo}
\dot S^{\rm (info. mot.)}=\Delta_i {\mathcal S}_{\rm fb}/\delta t=\dot I-J\Delta E \ge 0.
\end{equation}

\begin{figure}[t]
\centering
\[
\includegraphics[trim=.7cm .8cm .8cm 1.1cm, clip=true, scale=.36]{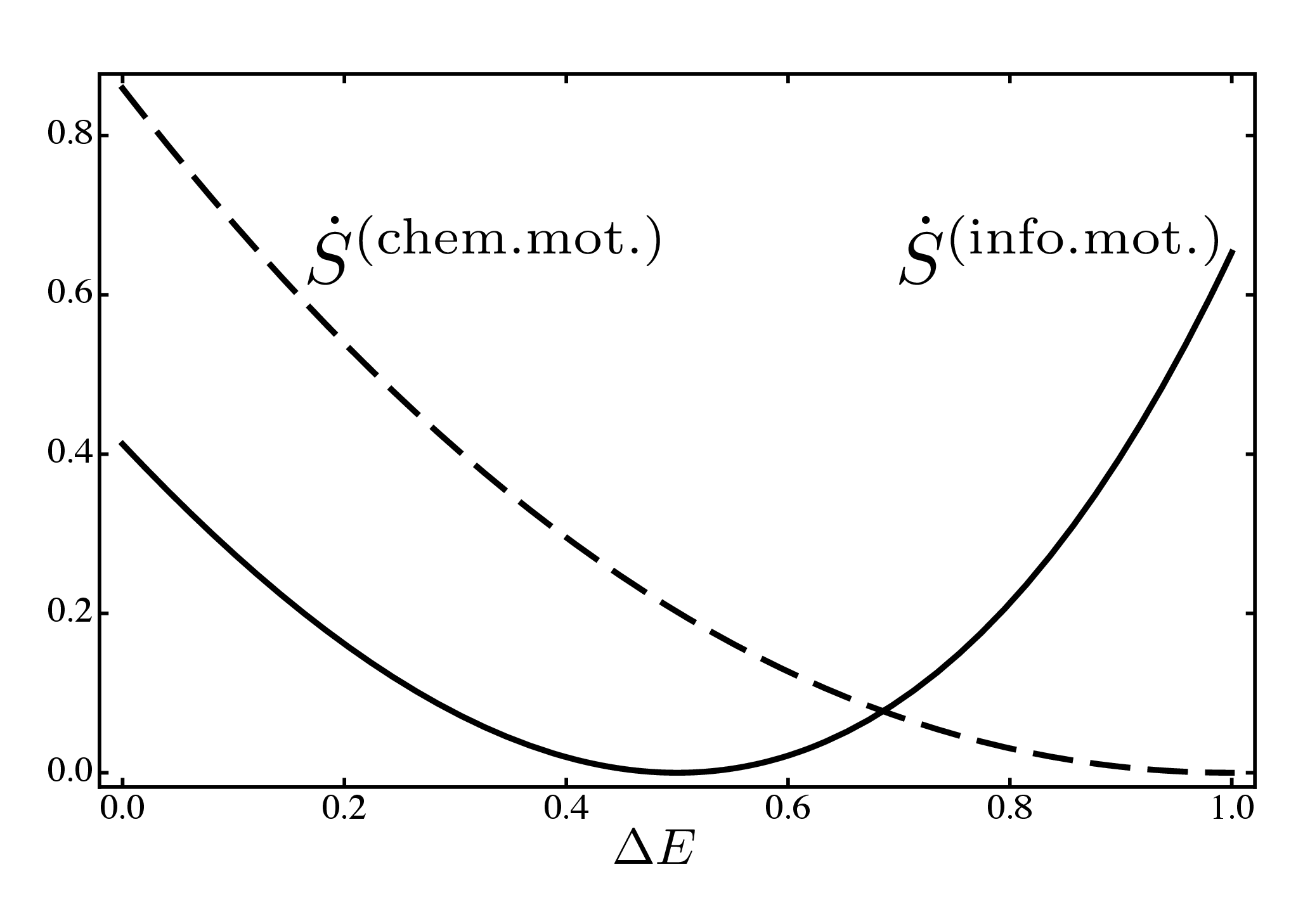}
\]
\caption{Plot of the entropy production rates for the chemical motor, $\dot S^{\rm (chem.mot.)}$ [Eq.~(\ref{schem})], and the information motor, $\dot S^{\rm (info.mot.)}$ [Eq.~(\ref{sinfo})], as  functions of the external force  $\Delta E$ for $q_{RL}=1$ and $\Delta\mu=\ln[(1-\epsilon)/\epsilon]=1$.
} \label{figentropies}
\end{figure}

In Fig.~\ref{figentropies},
we compare $\dot S^{\rm (chem.mot.)}$ [Eq.~(\ref{schem})] and $\dot S^{\rm (info.mot.)}$ [Eq.~(\ref{sinfo})] as functions of $\Delta E$.
The different switching mechanisms lead to qualitatively different thermodynamics, even though the dynamics are the same.
Most notably,  the chemical motor achieves the reversible limit only at the stall force $\Delta E=\Delta\mu $ when $J=0$, whereas the  information motor can operate with zero entropy production at a \emph{finite} current when $\Delta E=\Delta \mu/2$.
In this case, the feedback is reversible in the same spirit as other reversible controlled systems analyzed in Refs.~\cite{Horowitz2011,Horowitz:2011vg}.

To clarify the origin of this difference, we now analyze a physical realization of the information motor where the memory and  measurement  mechanism are included explicitly,  building on the mechanical Maxwell's demon introduced by Mandal and Jarzynski~\cite{Mandal:2012um}.

\begin{figure}[htb]
\begin{center}
\includegraphics[trim=.5cm 0cm .5cm 0cm, clip=true, scale=.19]{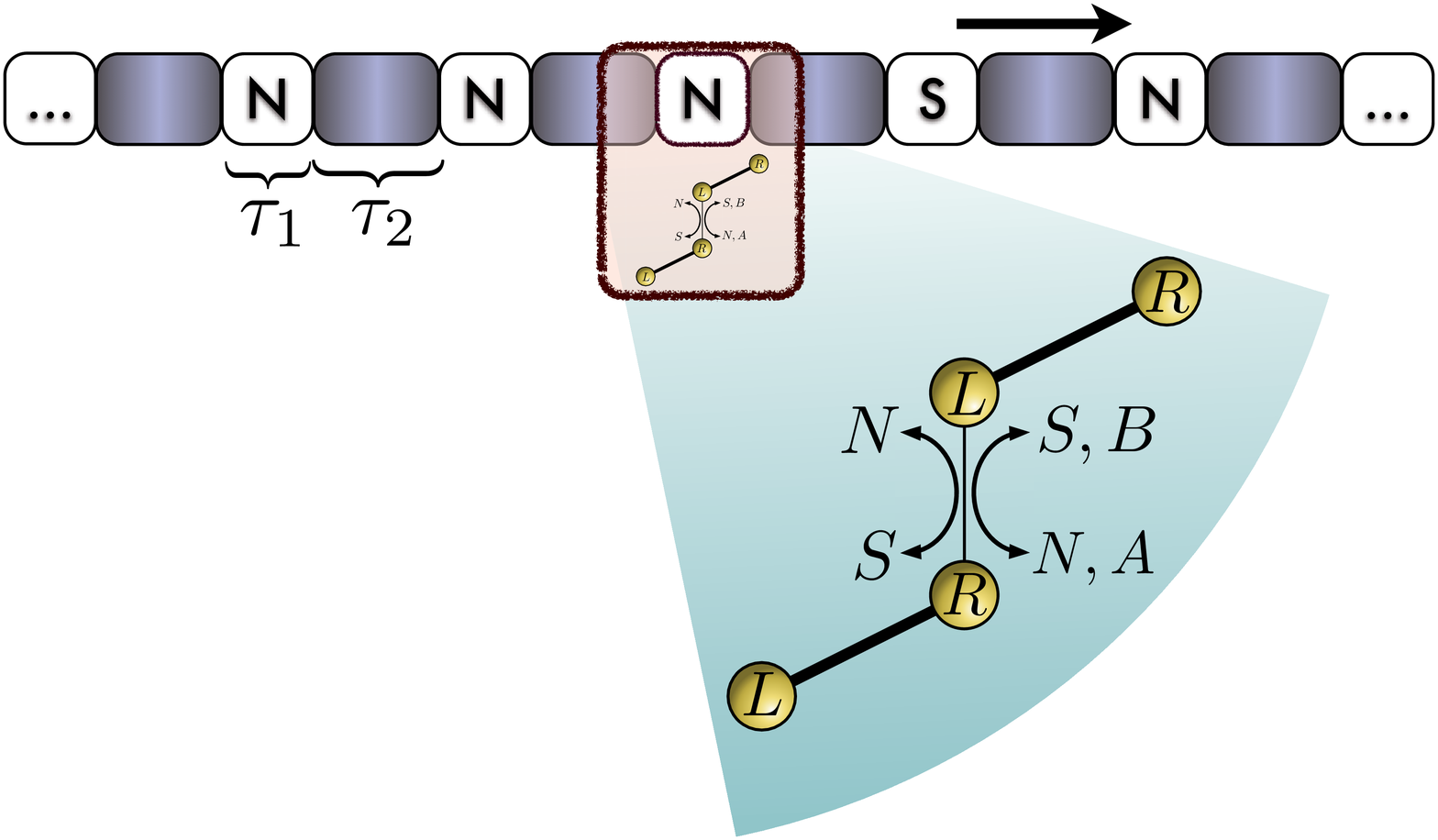}\\
\includegraphics[trim=0cm 0cm .4cm .6cm, clip=true, scale=.21]{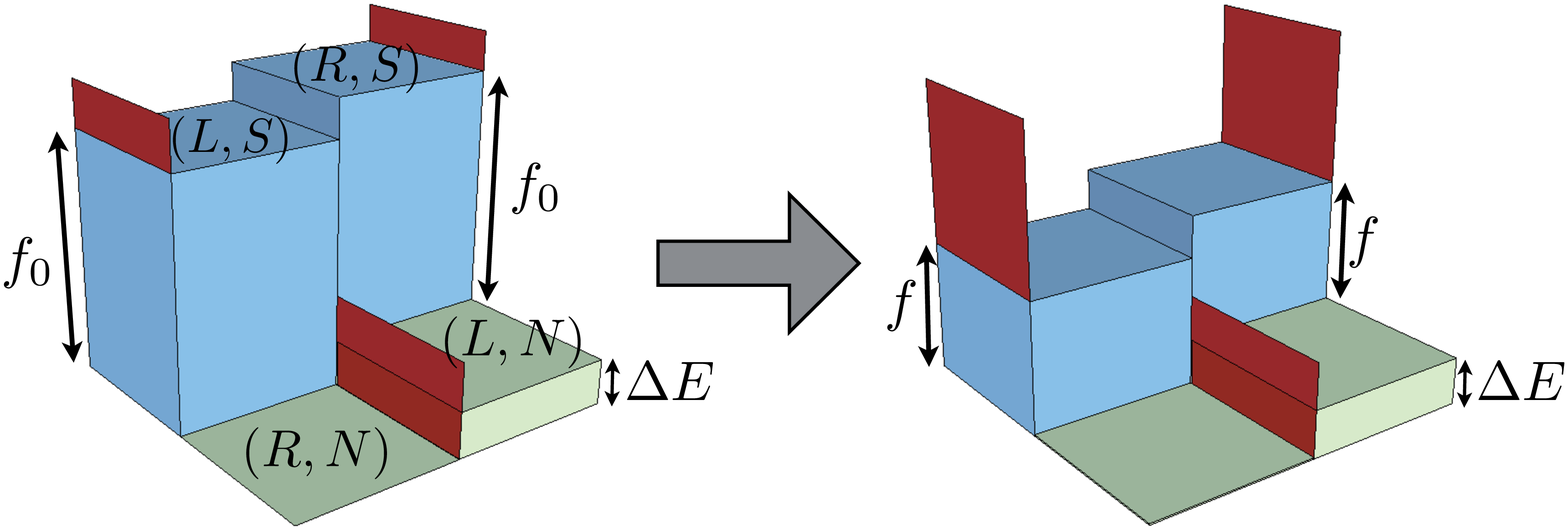}
\end{center}
\caption{Illustration of the information motor fed by a tape of two-state ($N$ and $S$) cells each initially in $N$ and separated by a time interval $\tau_2$ (upper figure).  Each cell couples to the ratchet for a duration $\tau_1$ during which $F_S$ is quasistatically lowered  from $f_0\to\infty$ to $f$ while the reactions $R+S\leftrightarrow L+N$ and $R+N+A \leftrightarrow L+S+B$ individually evolve in a time-dependent free energy landscape (lower figure). }
\label{figinfomotor}
\end{figure}

We model the memory as a tape composed of a series of two-state cells (or bits), with states $m=N,S$, and  free energies $F_N=0$ and $F_S=f_0 \to \infty$.
Initially, each cell is in $N$ -- which is  equilibrium ($p_N=1$, $p_S=0$).
 One at a time, each cell couples to the motor for a duration $\tau_1$, short compared to the diffusion ($r\tau_1\ll1$), through a  fast reaction that induces potential switches according to the scheme in Fig.~\ref{figinfomotor} at a rate of order $\gamma\gg r$.
We bias the $R\to L$ potential switches by mediating them with the same  out-of-equilibrium chemicals, $A$ and $B$, used in the chemical motor with $\Delta \mu=\ln(q_{LR}/q_{RL})$ through the formula $R+N+A \leftrightarrow L+S+B$.
Furthermore, we lower the free energy of $S$ quasistatically using the protocol $F_S(t)$ from $F_S(0)=f_0$ to $F_{S}(\tau_1)=f\equiv-\ln (q_{RL}\delta t)\gg 1$.
 Thus, even though the diffusion is frozen during $\tau_1$, each of the reactions  $R+S\leftrightarrow L+N$ and $R+N+A \leftrightarrow L+S+B$  independently evolve through a sequence of equilibrium states.
The dynamics during this period follows the master equation~\cite{VanKampen}
\begin{equation}\label{eq:master2}
\begin{split}
\dot{p}_{R,N}(t) &=\gamma p_{L,S}(t)-\gamma e^{F_N-F_{S}(t)+\Delta\mu}p_{R,N}(t)\\
\dot{p}_{L,N}(t) &=\gamma p_{R,S}(t)-\gamma e^{F_N-F_{S}(t)}p_{L,N}(t),
\end{split}
\end{equation}
with $p_{L,S}(t)=p_Rp_N-p_{R,N}(t)$ and $p_{R,S}(t)=p_Lp_N-p_{L,N}(t)$.
In the limit  $\gamma\tau_1\gg1$, we reproduce dynamically the  same correlations  as before, which may be verified by comparing the solution at $\tau_1$
\begin{equation}\label{ppost}
\left(\begin{array}{c}
p'_{R,N} \\
p'_{L,N} \\
\end{array}
\right)
=
\left(\begin{array}{c}
p_R[1-\alpha (1-\epsilon)\delta t] \\
p_L(1-\alpha\epsilon\delta t) \\
\end{array}
\right)
=
\left(\begin{array}{c}
p_R(1-q_{LR}\delta t) \\
p_L(1-q_{RL}\delta t) \\
\end{array}
\right),
\end{equation}
to the density of the information motor after the demon has acted~\cite{SuppMat}.
 Next, the cell decouples, and the motor relaxes to equilibrium by spatial diffusion for a time $\tau_2\gg1/r$ such that $\delta t=\tau_1+\tau_2$ is the total cycle time.
The various time scales can be summarized as $\gamma^{-1} \ll \tau_1 \sim {\dot F}_s^{-1}  \ll r^{-1} \ll \tau_2 \ll \alpha^{-1}$.

The thermodynamic analysis of each $\delta t$-cycle naturally decomposes  into two steps: the establishment of correlations, or measurement, during $\tau_1$, and the spatial relaxation during $\tau_2$ when the correlations are converted into work.

During $\tau_1$,  work is done by the $A\leftrightarrow B$ reaction, $W_{\rm chem}=p_{L,S}^\prime \Delta \mu$ [Eq.~(\ref{ppost})], and by quasistatically lowering $F_S$ from $f_0\to\infty$ to $f$, which to order $\delta t$ is
\begin{equation}
W_{\rm lower} = \int_{f_0}^{f}\left[\frac{p_R}{1+e^{f^\prime-\Delta\mu}}+\frac{p_L}{1+e^{f^\prime}}\right] df^\prime \simeq -p_S^\prime, 
\label{wshift}
\end{equation}
where $p_S^\prime=q_S\delta t=p'_{R,S}+p'_{L,S}$.
Within the nonequilibrium free energy framework, this work is interpreted as being used to form correlations $I(M^\prime,Z)$ [Eq.~(\ref{eq:infoRate})] while changing the memory's nonequilibrium free energy from  ${\mathcal F}(M)=0$ ($p_N=1$) by $\Delta{\mathcal F}(M)= {\mathcal F}(M^\prime) = p_S^\prime f-h(p_S^\prime)$, where $h$ is the binary Shannon entropy~\cite{Cover}. 
Inserting these expressions into Eq.~(\ref{eq:entMeas}), reveals, after a cumbersome  though straightforward algebraic manipulation, that
\begin{equation}\label{wmeas1}
\Delta_i {\mathcal S}_{\rm meas}=W_{\rm lower} + W_{\rm chem}- \Delta{\mathcal F}(M)-{I}(M^\prime,Z)=0,
\end{equation}
saturating the bound in Eq.~(\ref{eq:entMeas}).
Our protocol is reversible, because the initially equilibrium memory ($f_0\to\infty$, $p_S=0$) couples to an equilibrium ratchet, followed by a quasistatic isothermal shift in $F_S$.

The cycle is completed as the motor relaxes.
$I$ is converted into work $W_{\rm ext} = J\delta t\Delta E$ through a decrease of the nonequilibrium free energy to ${\mathcal F}(M^{\prime\prime},Z^{\prime\prime})={\mathcal F}(M^\prime)+{\mathcal F}(Z)$.
The resulting entropy production from Eq.~(\ref{eq:SinfoF}) is $\Delta_i {\mathcal S}_{\rm diff}=I(M^\prime,Z)-J\delta t\Delta E\ge 0$, reproducing 
per cycle the entropy production rate of the information motor $\Delta_i {\mathcal S}_{\rm diff}/\delta t = \dot S^{\rm (info. mot.)}$.
Thus, the total entropy production per cycle is $\Delta_i {\mathcal S}_{\rm tot}=\Delta_i {\mathcal S}_{\rm meas}+\Delta_i {\mathcal S}_{\rm diff} = \dot S^{\rm (info. mot.)} \delta t$, proving that we have a nonautonomous model for the information motor without explicit feedback that has the same dynamics \emph{and} thermodynamics.

Contact can be made with a traditional statement of the second law if we complete the cycle by restoring the memory to its initial configuration~\cite{Leff,Maruyama2009}. 
From Eq.~(\ref{eq:SinfoF}),  this requires a minimum work $W_{\rm rest}=-\Delta {\mathcal F}(M)$.
Using this minimum, the total work is $W_{\rm tot} = W_{\rm lower} + W_{\rm chem} + W_{\rm rest}-W_{\rm ext}=I-W_{\rm ext}$, by virtue of Eq.~({\ref{wmeas1}}).
Thus, plotted in Fig.~\ref{figentropies} is simply  $\dot S^{\rm (info.mot.)}={\dot W}_{\rm tot}$, the total energy dissipated in this cyclic isothermal process.

From this analysis, we conclude that the disparity between $\dot S^{\rm (info.mot.)}$ and $\dot S^{\rm (chem.mot.)}$ originates in the \emph{rapid and reversible} measurements that allow potential flips in the information motor to occur with zero entropy production, unlike in the chemical motor where flips produce entropy.
Such reversible measurements are possible due to the time-scale separation between the tape's internal transitions and the current ($\gamma \gg \alpha$), which allows the measurement (and flip) to be implemented using a \emph{nonautonomous reversible process}.
This mechanism allows for reversible transport --  a nonzero current without entropy production --  and can be regarded as an adiabatic pump~\cite{Sinitsyn2007,Parrondo1998,Rahav2008,Horowitz2009}.
Subsequently resetting the memory does not alter the information motor's entropy production, since it can always be accomplished reversibly.

In summary, we have examined an explicit physical mechanism that stores information in a memory to be used later.
This mechanism relies on the two-step interaction mediated by the tape that creates long-lived correlations. 
The sequential structure of the tape, however, seems less important: a reservoir of molecules $N$ and $S$ would produce the same behavior, as long as there was a mechanism establishing correlations for fixed intervals.
Remarkably, similar long-lived complexes are common in biology: 
they can be observed in molecular motors~\cite{Little2011}, enzymatic catalysis~\cite{Yang2011}, and sensory adaption~\cite{Lan2012}.  
It would be interesting to check if such complexes serve as a free energy storage and to uncover their role in information processing.

This work is funded by  Grants MOSAICO and ENFASIS (Spanish Government),  and MODELICO (Comunidad Autonoma de Madrid).
JMH is supported financially by the National Science Foundation (USA) International Research Fellowship under Grant No.~OISE-1059438. TS is supported financially by the Ministry of Education, Culture, Sports, Science and Technology of Japan (KAKENHI 11025807).

\bibliographystyle{apsrev4-1.bst} 
\bibliography{refs,PhysicsTexts}

\begin{thebibliography}{45}%
\makeatletter
\providecommand \@ifxundefined [1]{%
 \@ifx{#1\undefined}
}%
\providecommand \@ifnum [1]{%
 \ifnum #1\expandafter \@firstoftwo
 \else \expandafter \@secondoftwo
 \fi
}%
\providecommand \@ifx [1]{%
 \ifx #1\expandafter \@firstoftwo
 \else \expandafter \@secondoftwo
 \fi
}%
\providecommand \natexlab [1]{#1}%
\providecommand \enquote  [1]{``#1''}%
\providecommand \bibnamefont  [1]{#1}%
\providecommand \bibfnamefont [1]{#1}%
\providecommand \citenamefont [1]{#1}%
\providecommand \href@noop [0]{\@secondoftwo}%
\providecommand \href [0]{\begingroup \@sanitize@url \@href}%
\providecommand \@href[1]{\@@startlink{#1}\@@href}%
\providecommand \@@href[1]{\endgroup#1\@@endlink}%
\providecommand \@sanitize@url [0]{\catcode `\\12\catcode `\$12\catcode
  `\&12\catcode `\#12\catcode `\^12\catcode `\_12\catcode `\%12\relax}%
\providecommand \@@startlink[1]{}%
\providecommand \@@endlink[0]{}%
\providecommand \url  [0]{\begingroup\@sanitize@url \@url }%
\providecommand \@url [1]{\endgroup\@href {#1}{\urlprefix }}%
\providecommand \urlprefix  [0]{URL }%
\providecommand \Eprint [0]{\href }%
\@ifxundefined \urlstyle {%
  \providecommand \doi  [0]{\begingroup \@sanitize@url \@doi}%
  \providecommand \@doi [1]{\endgroup \@@startlink {\doibase
  #1}doi:\discretionary {}{}{}#1\@@endlink }%
}{%
  \providecommand \doi  [0]{doi:\discretionary{}{}{}\begingroup
  \urlstyle{rm}\Url }%
}%
\providecommand \doibase [0]{http://dx.doi.org/}%
\providecommand \Doi [0]{\begingroup \@sanitize@url \@Doi }%
\providecommand \@Doi  [1]{\endgroup\@@startlink{\doibase#1}\@@Doi}%
\providecommand \@@Doi [1]{#1\@@endlink}%
\providecommand \selectlanguage [0]{\@gobble}%
\providecommand \bibinfo  [0]{\@secondoftwo}%
\providecommand \bibfield  [0]{\@secondoftwo}%
\providecommand \translation [1]{[#1]}%
\providecommand \BibitemOpen [0]{}%
\providecommand \bibitemStop [0]{}%
\providecommand \bibitemNoStop [0]{.\EOS\space}%
\providecommand \EOS [0]{\spacefactor3000\relax}%
\providecommand \BibitemShut  [1]{\csname bibitem#1\endcsname}%
\bibitem [{\citenamefont {Landauer}(1991)}]{Landauer1991}%
  \BibitemOpen
  \bibfield  {author} {\bibinfo {author} {\bibfnamefont {R.}~\bibnamefont
  {Landauer}},\ }\href@noop {} {\bibfield  {journal} {\bibinfo  {journal}
  {Phys. Today},\ }\textbf {\bibinfo {volume} {44}},\ \bibinfo {pages} {23}
  (\bibinfo {year} {1991})}\BibitemShut {NoStop}%
\bibitem [{\citenamefont {Leff}\ and\ \citenamefont {Rex}(1990)}]{Leff}%
  \BibitemOpen
  \bibinfo {editor} {\bibfnamefont {H.~S.}\ \bibnamefont {Leff}}\ and\ \bibinfo
  {editor} {\bibfnamefont {A.~F.}\ \bibnamefont {Rex}},\ eds.,\ \href@noop {}
  {\emph {\bibinfo {title} {Maxwell's Demon: Entropy, Information,
  Computing}}}\ (\bibinfo  {publisher} {Princeton University Press, New
  Jersey},\ \bibinfo {year} {1990})\BibitemShut {NoStop}%
\bibitem [{\citenamefont {Maruyama}\ \emph {et~al.}(2009)\citenamefont
  {Maruyama}, \citenamefont {Nori},\ and\ \citenamefont
  {Vedral}}]{Maruyama2009}%
  \BibitemOpen
  \bibfield  {author} {\bibinfo {author} {\bibfnamefont {K.}~\bibnamefont
  {Maruyama}}, \bibinfo {author} {\bibfnamefont {F.}~\bibnamefont {Nori}}, \
  and\ \bibinfo {author} {\bibfnamefont {V.}~\bibnamefont {Vedral}},\
  }\href@noop {} {\bibfield  {journal} {\bibinfo  {journal} {Rev. Mod. Phys.},\
  }\textbf {\bibinfo {volume} {81}},\ \bibinfo {pages} {1} (\bibinfo {year}
  {2009})}\BibitemShut {NoStop}%
\bibitem [{\citenamefont {B\'erut}\ \emph {et~al.}(2011)\citenamefont
  {B\'erut}, \citenamefont {Arakelyan}, \citenamefont {Petrosyan},
  \citenamefont {Ciliberto}, \citenamefont {Dillenschneider},\ and\
  \citenamefont {Lutz}}]{Berut2011}%
  \BibitemOpen
  \bibfield  {author} {\bibinfo {author} {\bibfnamefont {A.}~\bibnamefont
  {B\'erut}}, \bibinfo {author} {\bibfnamefont {A.}~\bibnamefont {Arakelyan}},
  \bibinfo {author} {\bibfnamefont {A.}~\bibnamefont {Petrosyan}}, \bibinfo
  {author} {\bibfnamefont {S.}~\bibnamefont {Ciliberto}}, \bibinfo {author}
  {\bibfnamefont {R.}~\bibnamefont {Dillenschneider}}, \ and\ \bibinfo {author}
  {\bibfnamefont {E.}~\bibnamefont {Lutz}},\ }\href@noop {} {\bibfield
  {journal} {\bibinfo  {journal} {Nature},\ }\textbf {\bibinfo {volume}
  {483}},\ \bibinfo {pages} {187} (\bibinfo {year} {2011})}\BibitemShut
  {NoStop}%
\bibitem [{\citenamefont {Granger}\ and\ \citenamefont
  {Kantz}(2011)}]{Granger2011}%
  \BibitemOpen
  \bibfield  {author} {\bibinfo {author} {\bibfnamefont {L.}~\bibnamefont
  {Granger}}\ and\ \bibinfo {author} {\bibfnamefont {H.}~\bibnamefont
  {Kantz}},\ }\href@noop {} {\bibfield  {journal} {\bibinfo  {journal} {Phys.
  Rev. E},\ }\textbf {\bibinfo {volume} {84}},\ \bibinfo {pages} {061110}
  (\bibinfo {year} {2011})}\BibitemShut {NoStop}%
\bibitem [{\citenamefont {Bennett}(1982)}]{Bennett:1982wx}%
  \BibitemOpen
  \bibfield  {author} {\bibinfo {author} {\bibfnamefont {C.}~\bibnamefont
  {Bennett}},\ }\href@noop {} {\bibfield  {journal} {\bibinfo  {journal} {Int.
  J. Theor. Phys.},\ }\textbf {\bibinfo {volume} {21}},\ \bibinfo {pages} {905}
  (\bibinfo {year} {1982})}\BibitemShut {NoStop}%
\bibitem [{\citenamefont {Piechocinska}(2000)}]{Piechocinska2000}%
  \BibitemOpen
  \bibfield  {author} {\bibinfo {author} {\bibfnamefont {B.}~\bibnamefont
  {Piechocinska}},\ }\href@noop {} {\bibfield  {journal} {\bibinfo  {journal}
  {Phys. Rev. A},\ }\textbf {\bibinfo {volume} {61}},\ \bibinfo {pages}
  {062314} (\bibinfo {year} {2000})}\BibitemShut {NoStop}%
\bibitem [{\citenamefont {Landauer}(1986)}]{Landauer:2006gs}%
  \BibitemOpen
  \bibfield  {author} {\bibinfo {author} {\bibfnamefont {R.}~\bibnamefont
  {Landauer}},\ }\href@noop {} {\bibfield  {journal} {\bibinfo  {journal}
  {Phys. Scr.},\ }\textbf {\bibinfo {volume} {35}},\ \bibinfo {pages} {88}
  (\bibinfo {year} {1986})}\BibitemShut {NoStop}%
\bibitem [{\citenamefont {Dillenschneider}\ and\ \citenamefont
  {Lutz}(2009)}]{Dillenschneider2009}%
  \BibitemOpen
  \bibfield  {author} {\bibinfo {author} {\bibfnamefont {R.}~\bibnamefont
  {Dillenschneider}}\ and\ \bibinfo {author} {\bibfnamefont {E.}~\bibnamefont
  {Lutz}},\ }\href@noop {} {\bibfield  {journal} {\bibinfo  {journal} {Phys.
  Rev. Lett.},\ }\textbf {\bibinfo {volume} {102}},\ \bibinfo {pages} {210601}
  (\bibinfo {year} {2009})}\BibitemShut {NoStop}%
\bibitem [{\citenamefont {del Rio}\ \emph {et~al.}(2010)\citenamefont {del
  Rio}, \citenamefont {Aberg}, \citenamefont {Renner}, \citenamefont
  {Dhalsten},\ and\ \citenamefont {Vedral}}]{delRio2010}%
  \BibitemOpen
  \bibfield  {author} {\bibinfo {author} {\bibfnamefont {L.}~\bibnamefont {del
  Rio}}, \bibinfo {author} {\bibfnamefont {J.}~\bibnamefont {Aberg}}, \bibinfo
  {author} {\bibfnamefont {R.}~\bibnamefont {Renner}}, \bibinfo {author}
  {\bibfnamefont {O.}~\bibnamefont {Dhalsten}}, \ and\ \bibinfo {author}
  {\bibfnamefont {V.}~\bibnamefont {Vedral}},\ }\href@noop {} {\bibfield
  {journal} {\bibinfo  {journal} {Nature},\ }\textbf {\bibinfo {volume}
  {474}},\ \bibinfo {pages} {61} (\bibinfo {year} {2010})}\BibitemShut
  {NoStop}%
\bibitem [{\citenamefont {Sagawa}\ and\ \citenamefont
  {Ueda}(2009)}]{Sagawa2009}%
  \BibitemOpen
  \bibfield  {author} {\bibinfo {author} {\bibfnamefont {T.}~\bibnamefont
  {Sagawa}}\ and\ \bibinfo {author} {\bibfnamefont {M.}~\bibnamefont {Ueda}},\
  }\href@noop {} {\bibfield  {journal} {\bibinfo  {journal} {Phys. Rev.
  Lett.},\ }\textbf {\bibinfo {volume} {102}},\ \bibinfo {pages} {250602}
  (\bibinfo {year} {2009})}\BibitemShut {NoStop}%
\bibitem [{\citenamefont {Bennett}(1979)}]{Bennett:1979tb}%
  \BibitemOpen
  \bibfield  {author} {\bibinfo {author} {\bibfnamefont {C.}~\bibnamefont
  {Bennett}},\ }\href@noop {} {\bibfield  {journal} {\bibinfo  {journal}
  {BioSystems},\ }\textbf {\bibinfo {volume} {11}},\ \bibinfo {pages} {85}
  (\bibinfo {year} {1979})}\BibitemShut {NoStop}%
\bibitem [{\citenamefont {Andrieux}\ and\ \citenamefont
  {Gaspard}(2008)}]{Andrieux:2008tp}%
  \BibitemOpen
  \bibfield  {author} {\bibinfo {author} {\bibfnamefont {D.}~\bibnamefont
  {Andrieux}}\ and\ \bibinfo {author} {\bibfnamefont {P.}~\bibnamefont
  {Gaspard}},\ }\href@noop {} {\bibfield  {journal} {\bibinfo  {journal} {Proc.
  Nat. Ac. Sci.},\ }\textbf {\bibinfo {volume} {105}},\ \bibinfo {pages} {9516}
  (\bibinfo {year} {2008})}\BibitemShut {NoStop}%
\bibitem [{\citenamefont {Cao}\ \emph {et~al.}(2004)\citenamefont {Cao},
  \citenamefont {Dinis},\ and\ \citenamefont {Parrondo}}]{Cao2004}%
  \BibitemOpen
  \bibfield  {author} {\bibinfo {author} {\bibfnamefont {F.~J.}\ \bibnamefont
  {Cao}}, \bibinfo {author} {\bibfnamefont {L.}~\bibnamefont {Dinis}}, \ and\
  \bibinfo {author} {\bibfnamefont {J.~M.~R.}\ \bibnamefont {Parrondo}},\
  }\href@noop {} {\bibfield  {journal} {\bibinfo  {journal} {Phys. Rev.
  Lett.},\ }\textbf {\bibinfo {volume} {93}},\ \bibinfo {pages} {040603}
  (\bibinfo {year} {2004})}\BibitemShut {NoStop}%
\bibitem [{\citenamefont {Suzuki}\ and\ \citenamefont
  {Fujitani}(2009)}]{Suzuki2009}%
  \BibitemOpen
  \bibfield  {author} {\bibinfo {author} {\bibfnamefont {H.}~\bibnamefont
  {Suzuki}}\ and\ \bibinfo {author} {\bibfnamefont {Y.}~\bibnamefont
  {Fujitani}},\ }\href@noop {} {\bibfield  {journal} {\bibinfo  {journal} {J.
  Phys. Soc. Jap.},\ }\textbf {\bibinfo {volume} {78}},\ \bibinfo {pages}
  {074007} (\bibinfo {year} {2009})}\BibitemShut {NoStop}%
\bibitem [{\citenamefont {Vaikuntanathan}\ and\ \citenamefont
  {Jarzynski}(2011)}]{Vaikuntanathan2011}%
  \BibitemOpen
  \bibfield  {author} {\bibinfo {author} {\bibfnamefont {S.}~\bibnamefont
  {Vaikuntanathan}}\ and\ \bibinfo {author} {\bibfnamefont {C.}~\bibnamefont
  {Jarzynski}},\ }\href@noop {} {\bibfield  {journal} {\bibinfo  {journal}
  {Phys. Rev. E},\ }\textbf {\bibinfo {volume} {83}},\ \bibinfo {pages}
  {061120} (\bibinfo {year} {2011})}\BibitemShut {NoStop}%
\bibitem [{\citenamefont {Abreu}\ and\ \citenamefont
  {Seifert}(2011)}]{Abreu2011}%
  \BibitemOpen
  \bibfield  {author} {\bibinfo {author} {\bibfnamefont {D.}~\bibnamefont
  {Abreu}}\ and\ \bibinfo {author} {\bibfnamefont {U.}~\bibnamefont
  {Seifert}},\ }\href@noop {} {\bibfield  {journal} {\bibinfo  {journal}
  {Europhys. Lett.},\ }\textbf {\bibinfo {volume} {94}},\ \bibinfo {pages}
  {10001} (\bibinfo {year} {2011})}\BibitemShut {NoStop}%
\bibitem [{\citenamefont {Bauer}\ \emph {et~al.}(2012)\citenamefont {Bauer},
  \citenamefont {Abreu},\ and\ \citenamefont {Seifert}}]{Bauer2012}%
  \BibitemOpen
  \bibfield  {author} {\bibinfo {author} {\bibfnamefont {M.}~\bibnamefont
  {Bauer}}, \bibinfo {author} {\bibfnamefont {D.}~\bibnamefont {Abreu}}, \ and\
  \bibinfo {author} {\bibfnamefont {U.}~\bibnamefont {Seifert}},\ }\href@noop
  {} {\bibfield  {journal} {\bibinfo  {journal} {J. Phys. A: Math. Theor.},\
  }\textbf {\bibinfo {volume} {45}},\ \bibinfo {pages} {162001} (\bibinfo
  {year} {2012})}\BibitemShut {NoStop}%
\bibitem [{\citenamefont {Horowitz}\ and\ \citenamefont
  {Parrondo}(2011){\natexlab{a}}}]{Horowitz2011}%
  \BibitemOpen
  \bibfield  {author} {\bibinfo {author} {\bibfnamefont {J.~M.}\ \bibnamefont
  {Horowitz}}\ and\ \bibinfo {author} {\bibfnamefont {J.~M.~R.}\ \bibnamefont
  {Parrondo}},\ }\href@noop {} {\bibfield  {journal} {\bibinfo  {journal}
  {Europhys. Lett.},\ }\textbf {\bibinfo {volume} {95}},\ \bibinfo {pages}
  {10005} (\bibinfo {year} {2011}{\natexlab{a}})}\BibitemShut {NoStop}%
\bibitem [{\citenamefont {Horowitz}\ and\ \citenamefont
  {Parrondo}(2011){\natexlab{b}}}]{Horowitz:2011vg}%
  \BibitemOpen
  \bibfield  {author} {\bibinfo {author} {\bibfnamefont {J.~M.}\ \bibnamefont
  {Horowitz}}\ and\ \bibinfo {author} {\bibfnamefont {J.~M.~R.}\ \bibnamefont
  {Parrondo}},\ }\href@noop {} {\bibfield  {journal} {\bibinfo  {journal} {New
  J. Phys.},\ }\textbf {\bibinfo {volume} {13}},\ \bibinfo {pages} {123019}
  (\bibinfo {year} {2011}{\natexlab{b}})}\BibitemShut {NoStop}%
\bibitem [{\citenamefont {Lopez}\ \emph {et~al.}(2008)\citenamefont {Lopez},
  \citenamefont {Kuwada}, \citenamefont {Craig}, \citenamefont {Long},\ and\
  \citenamefont {Linke}}]{Lopez2008}%
  \BibitemOpen
  \bibfield  {author} {\bibinfo {author} {\bibfnamefont {B.~J.}\ \bibnamefont
  {Lopez}}, \bibinfo {author} {\bibfnamefont {N.~J.}\ \bibnamefont {Kuwada}},
  \bibinfo {author} {\bibfnamefont {E.~M.}\ \bibnamefont {Craig}}, \bibinfo
  {author} {\bibfnamefont {B.~R.}\ \bibnamefont {Long}}, \ and\ \bibinfo
  {author} {\bibfnamefont {H.}~\bibnamefont {Linke}},\ }\href@noop {}
  {\bibfield  {journal} {\bibinfo  {journal} {Phys. Rev. Lett.},\ }\textbf
  {\bibinfo {volume} {101}},\ \bibinfo {pages} {220601} (\bibinfo {year}
  {2008})}\BibitemShut {NoStop}%
\bibitem [{\citenamefont {Toyabe}\ \emph {et~al.}(2010)\citenamefont {Toyabe},
  \citenamefont {Sagawa}, \citenamefont {Ueda}, \citenamefont {Muneyuki},\ and\
  \citenamefont {Sano}}]{Toyabe2010}%
  \BibitemOpen
  \bibfield  {author} {\bibinfo {author} {\bibfnamefont {S.}~\bibnamefont
  {Toyabe}}, \bibinfo {author} {\bibfnamefont {T.}~\bibnamefont {Sagawa}},
  \bibinfo {author} {\bibfnamefont {M.}~\bibnamefont {Ueda}}, \bibinfo {author}
  {\bibfnamefont {E.}~\bibnamefont {Muneyuki}}, \ and\ \bibinfo {author}
  {\bibfnamefont {M.}~\bibnamefont {Sano}},\ }\href@noop {} {\bibfield
  {journal} {\bibinfo  {journal} {Nature Phys.},\ }\textbf {\bibinfo {volume}
  {6}},\ \bibinfo {pages} {988} (\bibinfo {year} {2010})}\BibitemShut {NoStop}%
\bibitem [{\citenamefont {Parrondo}\ and\ \citenamefont
  {De~Cisneros}(2002)}]{Parrondo2002}%
  \BibitemOpen
  \bibfield  {author} {\bibinfo {author} {\bibfnamefont {J.~M.~R.}\
  \bibnamefont {Parrondo}}\ and\ \bibinfo {author} {\bibfnamefont {B.~J.}\
  \bibnamefont {De~Cisneros}},\ }\href@noop {} {\bibfield  {journal} {\bibinfo
  {journal} {Appl. Phys. A},\ }\textbf {\bibinfo {volume} {75}},\ \bibinfo
  {pages} {179} (\bibinfo {year} {2002})}\BibitemShut {NoStop}%
\bibitem [{\citenamefont {Reimann}(2002)}]{Reimann2002}%
  \BibitemOpen
  \bibfield  {author} {\bibinfo {author} {\bibfnamefont {P.}~\bibnamefont
  {Reimann}},\ }\href@noop {} {\bibfield  {journal} {\bibinfo  {journal} {Phys.
  Rep.},\ }\textbf {\bibinfo {volume} {361}},\ \bibinfo {pages} {67} (\bibinfo
  {year} {2002})}\BibitemShut {NoStop}%
\bibitem [{\citenamefont {Van~Kampen}(2007)}]{VanKampen}%
  \BibitemOpen
  \bibfield  {author} {\bibinfo {author} {\bibfnamefont {N.~G.}\ \bibnamefont
  {Van~Kampen}},\ }\href@noop {} {\emph {\bibinfo {title} {Stochastic Processes
  in Physics and Chemistry}}},\ \bibinfo {edition} {3rd}\ ed.\ (\bibinfo
  {publisher} {Elsevier Ltd., New York},\ \bibinfo {year} {2007})\BibitemShut
  {NoStop}%
\bibitem [{Sup()}]{SuppMat}%
  \BibitemOpen
  \href@noop {} {}\bibinfo {note} {See Supplemental Material}\BibitemShut
  {NoStop}%
\bibitem [{\citenamefont {Sagawa}\ and\ \citenamefont
  {Ueda}(2008)}]{Sagawa2008}%
  \BibitemOpen
  \bibfield  {author} {\bibinfo {author} {\bibfnamefont {T.}~\bibnamefont
  {Sagawa}}\ and\ \bibinfo {author} {\bibfnamefont {M.}~\bibnamefont {Ueda}},\
  }\href@noop {} {\bibfield  {journal} {\bibinfo  {journal} {Phys. Rev.
  Lett.},\ }\textbf {\bibinfo {volume} {100}},\ \bibinfo {pages} {080403}
  (\bibinfo {year} {2008})}\BibitemShut {NoStop}%
\bibitem [{\citenamefont {Sagawa}\ and\ \citenamefont
  {Ueda}(2010)}]{Sagawa2010}%
  \BibitemOpen
  \bibfield  {author} {\bibinfo {author} {\bibfnamefont {T.}~\bibnamefont
  {Sagawa}}\ and\ \bibinfo {author} {\bibfnamefont {M.}~\bibnamefont {Ueda}},\
  }\href@noop {} {\bibfield  {journal} {\bibinfo  {journal} {Phys. Rev.
  Lett.},\ }\textbf {\bibinfo {volume} {104}},\ \bibinfo {pages} {090602}
  (\bibinfo {year} {2010})}\BibitemShut {NoStop}%
\bibitem [{\citenamefont {Ponmurugan}(2010)}]{Ponmurugan2010}%
  \BibitemOpen
  \bibfield  {author} {\bibinfo {author} {\bibfnamefont {M.}~\bibnamefont
  {Ponmurugan}},\ }\Doi {10.1103/PhysRevE.82.031129} {\bibfield  {journal}
  {\bibinfo  {journal} {Phys. Rev. E},\ }\textbf {\bibinfo {volume} {82}},\
  \bibinfo {pages} {031129} (\bibinfo {year} {2010})}\BibitemShut {NoStop}%
\bibitem [{\citenamefont {Abreu}\ and\ \citenamefont
  {Seifert}(2012)}]{Abreu2012}%
  \BibitemOpen
  \bibfield  {author} {\bibinfo {author} {\bibfnamefont {D.}~\bibnamefont
  {Abreu}}\ and\ \bibinfo {author} {\bibfnamefont {U.}~\bibnamefont
  {Seifert}},\ }\href@noop {} {\bibfield  {journal} {\bibinfo  {journal} {Phys.
  Rev. Lett.},\ }\textbf {\bibinfo {volume} {108}},\ \bibinfo {pages} {030601}
  (\bibinfo {year} {2012})}\BibitemShut {NoStop}%
\bibitem [{\citenamefont {Horowitz}\ and\ \citenamefont
  {Vaikuntanathan}(2010)}]{Horowitz2010}%
  \BibitemOpen
  \bibfield  {author} {\bibinfo {author} {\bibfnamefont {J.~M.}\ \bibnamefont
  {Horowitz}}\ and\ \bibinfo {author} {\bibfnamefont {S.}~\bibnamefont
  {Vaikuntanathan}},\ }\href@noop {} {\bibfield  {journal} {\bibinfo  {journal}
  {Phys. Rev. E},\ }\textbf {\bibinfo {volume} {82}},\ \bibinfo {pages}
  {061120} (\bibinfo {year} {2010})}\BibitemShut {NoStop}%
\bibitem [{\citenamefont {Fujitani}\ and\ \citenamefont
  {Suzuki}(2010)}]{Suzuki2010}%
  \BibitemOpen
  \bibfield  {author} {\bibinfo {author} {\bibfnamefont {Y.}~\bibnamefont
  {Fujitani}}\ and\ \bibinfo {author} {\bibfnamefont {H.}~\bibnamefont
  {Suzuki}},\ }\href@noop {} {\bibfield  {journal} {\bibinfo  {journal} {J.
  Phys. Soc. Jap.},\ }\textbf {\bibinfo {volume} {79}},\ \bibinfo {pages}
  {104003} (\bibinfo {year} {2010})}\BibitemShut {NoStop}%
\bibitem [{\citenamefont {Sagawa}\ and\ \citenamefont
  {Ueda}(2012)}]{Sagawa2012}%
  \BibitemOpen
  \bibfield  {author} {\bibinfo {author} {\bibfnamefont {T.}~\bibnamefont
  {Sagawa}}\ and\ \bibinfo {author} {\bibfnamefont {M.}~\bibnamefont {Ueda}},\
  }\href@noop {} {\bibfield  {journal} {\bibinfo  {journal} {Phys. Rev.
  Lett.},\ }\textbf {\bibinfo {volume} {109}},\ \bibinfo {pages} {180602}
  (\bibinfo {year} {2012})}\BibitemShut {NoStop}%
\bibitem [{\citenamefont {Esposito}\ and\ \citenamefont {Van~den
  Broeck}(2011)}]{Esposito2011}%
  \BibitemOpen
  \bibfield  {author} {\bibinfo {author} {\bibfnamefont {M.}~\bibnamefont
  {Esposito}}\ and\ \bibinfo {author} {\bibfnamefont {C.}~\bibnamefont {Van~den
  Broeck}},\ }\href@noop {} {\bibfield  {journal} {\bibinfo  {journal}
  {Europhys. Lett.},\ }\textbf {\bibinfo {volume} {95}},\ \bibinfo {pages}
  {40004} (\bibinfo {year} {2011})}\BibitemShut {NoStop}%
\bibitem [{\citenamefont {Deffner}\ and\ \citenamefont
  {Lutz}(2012)}]{Deffner:2012tm}%
  \BibitemOpen
  \bibfield  {author} {\bibinfo {author} {\bibfnamefont {S.}~\bibnamefont
  {Deffner}}\ and\ \bibinfo {author} {\bibfnamefont {E.}~\bibnamefont {Lutz}},\
  }\href@noop {} {\bibfield  {journal} {\bibinfo  {journal} {arXiv:1201.3888}}
  (\bibinfo {year} {2012})}\BibitemShut {NoStop}%
\bibitem [{\citenamefont {Hill}(1977)}]{Hill}%
  \BibitemOpen
  \bibfield  {author} {\bibinfo {author} {\bibfnamefont {T.~L.}\ \bibnamefont
  {Hill}},\ }\href@noop {} {\emph {\bibinfo {title} {Free Energy Transduction
  in Biology}}}\ (\bibinfo  {publisher} {Academic Press, New York},\ \bibinfo
  {year} {1977})\BibitemShut {NoStop}%
\bibitem [{\citenamefont {Cover}\ and\ \citenamefont {Thomas}(2006)}]{Cover}%
  \BibitemOpen
  \bibfield  {author} {\bibinfo {author} {\bibfnamefont {T.~M.}\ \bibnamefont
  {Cover}}\ and\ \bibinfo {author} {\bibfnamefont {J.~A.}\ \bibnamefont
  {Thomas}},\ }\href@noop {} {\emph {\bibinfo {title} {Elements of Information
  Theory}}},\ \bibinfo {edition} {2nd}\ ed.\ (\bibinfo  {publisher}
  {Wiley-Interscience},\ \bibinfo {year} {2006})\BibitemShut {NoStop}%
\bibitem [{\citenamefont {Mandal}\ and\ \citenamefont
  {Jarzynski}(2012)}]{Mandal:2012um}%
  \BibitemOpen
  \bibfield  {author} {\bibinfo {author} {\bibfnamefont {D.}~\bibnamefont
  {Mandal}}\ and\ \bibinfo {author} {\bibfnamefont {C.}~\bibnamefont
  {Jarzynski}},\ }\href@noop {} {\bibfield  {journal} {\bibinfo  {journal}
  {Proc. Nat. Ac. Sci.}} (\bibinfo {year} {2012})}\BibitemShut {NoStop}%
\bibitem [{\citenamefont {Sinitsyn}\ and\ \citenamefont
  {Nemenman}(2007)}]{Sinitsyn2007}%
  \BibitemOpen
  \bibfield  {author} {\bibinfo {author} {\bibfnamefont {N.~A.}\ \bibnamefont
  {Sinitsyn}}\ and\ \bibinfo {author} {\bibfnamefont {I.}~\bibnamefont
  {Nemenman}},\ }\href@noop {} {\bibfield  {journal} {\bibinfo  {journal}
  {Europhys. Lett.},\ }\textbf {\bibinfo {volume} {77}},\ \bibinfo {pages}
  {58001} (\bibinfo {year} {2007})}\BibitemShut {NoStop}%
\bibitem [{\citenamefont {Parrondo}(1998)}]{Parrondo1998}%
  \BibitemOpen
  \bibfield  {author} {\bibinfo {author} {\bibfnamefont {J.~M.~R.}\
  \bibnamefont {Parrondo}},\ }\href@noop {} {\bibfield  {journal} {\bibinfo
  {journal} {Phys. Rev. E},\ }\textbf {\bibinfo {volume} {57}},\ \bibinfo
  {pages} {7297} (\bibinfo {year} {1998})}\BibitemShut {NoStop}%
\bibitem [{\citenamefont {Rahav}\ \emph {et~al.}(2008)\citenamefont {Rahav},
  \citenamefont {Horowitz},\ and\ \citenamefont {Jarzynski}}]{Rahav2008}%
  \BibitemOpen
  \bibfield  {author} {\bibinfo {author} {\bibfnamefont {S.}~\bibnamefont
  {Rahav}}, \bibinfo {author} {\bibfnamefont {J.}~\bibnamefont {Horowitz}}, \
  and\ \bibinfo {author} {\bibfnamefont {C.}~\bibnamefont {Jarzynski}},\
  }\href@noop {} {\bibfield  {journal} {\bibinfo  {journal} {Phys. Rev.
  Lett.},\ }\textbf {\bibinfo {volume} {101}},\ \bibinfo {pages} {140602}
  (\bibinfo {year} {2008})}\BibitemShut {NoStop}%
\bibitem [{\citenamefont {Horowitz}\ and\ \citenamefont
  {Jarzynski}(2009)}]{Horowitz2009}%
  \BibitemOpen
  \bibfield  {author} {\bibinfo {author} {\bibfnamefont {J.~M.}\ \bibnamefont
  {Horowitz}}\ and\ \bibinfo {author} {\bibfnamefont {C.}~\bibnamefont
  {Jarzynski}},\ }\href@noop {} {\bibfield  {journal} {\bibinfo  {journal} {J.
  Stat. Phys.},\ }\textbf {\bibinfo {volume} {136}},\ \bibinfo {pages} {917}
  (\bibinfo {year} {2009})}\BibitemShut {NoStop}%
\bibitem [{\citenamefont {Little}\ \emph {et~al.}(2011)\citenamefont {Little},
  \citenamefont {Steel}, \citenamefont {Bai}, \citenamefont {Sowa},
  \citenamefont {Bilyard}, \citenamefont {Mueller}, \citenamefont {Berry},\
  and\ \citenamefont {Jones}}]{Little2011}%
  \BibitemOpen
  \bibfield  {author} {\bibinfo {author} {\bibfnamefont {M.~A.}\ \bibnamefont
  {Little}}, \bibinfo {author} {\bibfnamefont {B.~C.}\ \bibnamefont {Steel}},
  \bibinfo {author} {\bibfnamefont {F.}~\bibnamefont {Bai}}, \bibinfo {author}
  {\bibfnamefont {Y.}~\bibnamefont {Sowa}}, \bibinfo {author} {\bibfnamefont
  {T.}~\bibnamefont {Bilyard}}, \bibinfo {author} {\bibfnamefont {D.~M.}\
  \bibnamefont {Mueller}}, \bibinfo {author} {\bibfnamefont {R.~M.}\
  \bibnamefont {Berry}}, \ and\ \bibinfo {author} {\bibfnamefont {N.~S.}\
  \bibnamefont {Jones}},\ }\href@noop {} {\bibfield  {journal} {\bibinfo
  {journal} {Biophys. J.},\ }\textbf {\bibinfo {volume} {101}},\ \bibinfo
  {pages} {477} (\bibinfo {year} {2011})}\BibitemShut {NoStop}%
\bibitem [{\citenamefont {Yang}\ \emph {et~al.}(2011)\citenamefont {Yang},
  \citenamefont {Cao}, \citenamefont {Silbey},\ and\ \citenamefont
  {Sung}}]{Yang2011}%
  \BibitemOpen
  \bibfield  {author} {\bibinfo {author} {\bibfnamefont {S.}~\bibnamefont
  {Yang}}, \bibinfo {author} {\bibfnamefont {J.}~\bibnamefont {Cao}}, \bibinfo
  {author} {\bibfnamefont {R.~J.}\ \bibnamefont {Silbey}}, \ and\ \bibinfo
  {author} {\bibfnamefont {J.}~\bibnamefont {Sung}},\ }\href@noop {} {\bibfield
   {journal} {\bibinfo  {journal} {Biophys. J.},\ }\textbf {\bibinfo {volume}
  {101}},\ \bibinfo {pages} {519} (\bibinfo {year} {2011})}\BibitemShut
  {NoStop}%
\bibitem [{\citenamefont {Lan}\ \emph {et~al.}(2012)\citenamefont {Lan},
  \citenamefont {Sartori}, \citenamefont {Neumann}, \citenamefont {Sourjik},\
  and\ \citenamefont {Tu}}]{Lan2012}%
  \BibitemOpen
  \bibfield  {author} {\bibinfo {author} {\bibfnamefont {G.}~\bibnamefont
  {Lan}}, \bibinfo {author} {\bibfnamefont {P.}~\bibnamefont {Sartori}},
  \bibinfo {author} {\bibfnamefont {S.}~\bibnamefont {Neumann}}, \bibinfo
  {author} {\bibfnamefont {V.}~\bibnamefont {Sourjik}}, \ and\ \bibinfo
  {author} {\bibfnamefont {Y.}~\bibnamefont {Tu}},\ }\href@noop {} {\bibfield
  {journal} {\bibinfo  {journal} {Nature Phys.},\ }\textbf {\bibinfo {volume}
  {8}},\ \bibinfo {pages} {422} (\bibinfo {year} {2012})}\BibitemShut {NoStop}%
\end{thebibliography}%

\setcounter{equation}{0}

\begin{widetext}
\section{Supplemental Material}
\subsection{Brownian ratchet thermodynamics}
\subsubsection{Derivation of Eq.~(2)}

The master equation (Eq.~(1) of the main text) is a continuity equation for the flow of probability, which may be made explicit in terms of the currents at time $t$,
\begin{equation}
J_{\rm diff}(t)=r_{RL}p_L (t) - r_{LR} p_R(t),
\end{equation}
representing the average net number of thermally activated diffusive jumps per unit time from $L\to R$  within each box, and 
\begin{equation}
J_{\rm chem}(t)=q_{LR} p_R (t)  - q_{RL} p_L (t),
\end{equation}
which is the average net number of transitions per unit time from $R\to L$ mediated by the chemical reaction or demon.
With this notation, Eq.~(1) takes the form
\begin{equation}\label{eq:mastCurr}
\begin{split}
{\dot p}_R &= J_{\rm diff}(t)-J_{\rm chem}(t)\\
{\dot p}_L &= J_{\rm chem}(t)-J_{\rm diff}(t),
\end{split}
\end{equation}
demonstrating that the change in the distribution of the ratchet is due to flows of probability, or current, in and out of each state.

As the Brownian ratchet evolves, work $\Delta E$ is extracted by the external force $F=2\Delta E/l$ every time the particle makes a diffusive jump up the slope from $L\to R$, whereas work $-\Delta E$ is extracted when the particle jumps down the slope.
As a result, the average rate of work extraction is~[23]
\begin{equation}\label{eq:work}
{\dot W}_{\rm ext}=J_{\rm diff}(t)\Delta E.
\end{equation}

In the stationary state ($\dot p_R (t) = \dot p_L (t) = 0$), Eq.~(\ref{eq:mastCurr}) above requires that the currents are equal $ J_{\rm diff}  = J_{\rm chem} \equiv J$.
Equation~(2) of the main text follows.

\subsubsection{Derivation of Eq.~(3)}

In the limit of $q \ll r$, Eq.~(1) of the main text reduces to
\begin{equation}
\begin{split}
\dot p_R (t) &= r_{RL} p_L (t) - r_{LR} p_R(t)=J_{\rm diff}(t), \\
\dot p_L (t) &= r_{LR} p_R (t)  - r_{RL} p_L (t)=-J_{\rm diff}(t).
\end{split}
\end{equation}
Its stationary solution ($\dot p_R (t) = \dot p_L (t) = 0$), obtained from a simple algebraic manipulation,  is  given in Eq.~(3) of the main text.
The stationary current remains nonzero due to potential switches; therefore it is given solely by the chemical current, which to lowest order in $q/r$ is
\begin{equation}
J=J_{\rm chem}=q_{LR} p_R  - q_{RL} p_L.
\end{equation}

\subsubsection{Derivation of Eq.~(4)}

In the chemical motor, the energy extracted by the force $F$ is provided by the free energy of the chemical reaction $A\leftrightarrow B$.
Every time the ratchet transitions and converts an $A$ ($B$) molecule into a $B$ ($A$) chemical free energy $\Delta \mu$ ($-\Delta \mu$) is consumed.
Therefore the average rate of free energy consumption is~[23]
\begin{equation}\label{eq:free}
\dot F_{\rm chem} (t) =  q_{LR} p_R (t) \Delta \mu +  q_{RL} p_L (t) (-\Delta \mu ) = J_{\rm chem} (t)  \Delta \mu.
\end{equation}

The entropy production rate is then expressed as~[23]
\begin{equation}\label{eq:Sdot}
\dot S^{\rm (chem. mot.)} (t) = \dot H_{\rm motor} + \dot F_{\rm chem} (t) - \dot W_{\rm ext} (t), 
\end{equation}
where
\begin{equation}
\dot H_{\rm motor} = - \frac{d}{dt} \sum_{x = L,R} p_x(t) \ln p_x(t)
\end{equation}
is the rate at which the Shannon entropy of the motor increases.

In the stationary state ($\dot p_R (t) = \dot p_L (t) = 0$), $J=J_{\rm chem}=J_{\rm diff}$ and ${\dot H}_{\rm motor}=0$.
Substituting these observations into Eqs.~(\ref{eq:work}), (\ref{eq:free}), and (\ref{eq:Sdot}) above, recovers Eq.~(4) of the main text. 

\subsection{Nonequilibrium free energy: derivation of Eq.~(7)}

We next derive Eq.~(7) in the main text.
During measurement the memory state changes $M\to M^\prime$, but the system state remains unchanged $Z=Z^\prime$.
Thus, after the measurement, we have from the definition of the nonequilibrium free energy in Eq.~(5) of the text
\begin{equation}\label{eq:freeNon}
\mathcal F (M', Z') =  \mathcal{F}(M', Z)= F(M', Z) - H(M',Z).
\end{equation}
The definition of mutual information in Eq.~(8) can be rearranged as 
\begin{equation}
H(M',Z)=H(M^\prime)+H(Z)-I(M^\prime,Z).
\end{equation}
Substituting this and the assumption of additive bare free energies $F(M^\prime,Z)=F(M^\prime)+F(Z)$ into Eq.~(\ref{eq:freeNon}) leads to
\begin{align}
\mathcal F (M', Z') &= F(M') - H(M^\prime) +F(Z) - H(Z) + I(M',Z) \\
&= {\mathcal F}(M') + {\mathcal F}(Z) + I(M',Z).
\end{align}

\subsection{Information rate: derivation of Eq.~(12)}

Let $Z = L, R$ be the pre-measurement state of the motor, and $M' = N,S$ be the post-measurement state of the memory.
The joint probability distribution of $(M',Z)$ is
\begin{equation}\label{eq:meas}
\left( 
\begin{array}{c}
p_{R,N} \\
p_{R,S} \\
p_{L,N} \\
p_{L,S}
\end{array}
\right) =  \left(
\begin{array}{c}
(1 - \alpha (1- \varepsilon ) \delta t) p_R  \\
\alpha (1- \varepsilon )\delta t p_R  \\
(1-\alpha \varepsilon \delta t) p_L \\
\alpha \varepsilon  \delta t p_L
\end{array}
\right) =  \left(
\begin{array}{c}
(1 - q_{LR}\delta t) p_R  \\
q_{LR} p_R \delta t \\
(1-q_{RL} \delta t) p_L \\
q_{RL} p_L  \delta t
\end{array}
\right).
\end{equation}
The mutual information $I(M',Z)$ is then be expressed as~[31,37]
\begin{equation}
\begin{split}
I(M',Z) =~&p_{R,N} \ln \frac{p_{R,N}}{p_R (p_{R,N} + p_{L,N})} +p_{L,N} \ln \frac{p_{L,N}}{p_L (p_{R,N} + p_{L,N})} \\
&+ p_{R,S}\ln \frac{p_{R,S}}{p_R (p_{R,S} + p_{L,S})} +p_{L,S} \ln \frac{p_{L,S}}{p_L (p_{R,S} + p_{L,S})} \\
=~& (1- q_{LR} \delta t) p_R \ln \frac{1-q_{LR}\delta t}{1 - ( q_{LR}p_R + q_{RL} p_L ) \delta t} \\
&+(1-q_{RL} \delta t) p_L  \ln \frac{1-q_{RL}\delta t}{1 - ( q_{LR}p_R + q_{RL} p_L ) \delta t} \\
&+ q_{LR}p_R \delta t \ln \frac{q_{LR}}{q_{LR}p_R + q_{RL} p_L} +q_{RL} p_L  \ln \frac{q_{RL}}{q_{LR}p_R + q_{RL} p_L}.
\end{split}
\end{equation}
The first and second lines cancel:
\begin{equation}
\begin{split}
&(1- q_{LR} \delta t) p_R \ln \frac{1-q_{LR}\delta t}{1 - ( q_{LR}p_R + q_{RL} p_L ) \delta t} +(1-q_{RL} \delta t) p_L  \ln \frac{1-q_{RL}\delta t}{1 - ( q_{LR}p_R + q_{RL} p_L ) \delta t} \\
&= (1- q_{LR} \delta t) p_R  (- q_{LR} + ( q_{LR}p_R + q_{RL} p_L ) )\delta t \\
&~~~+ (1-q_{RL} \delta t) p_L  (- q_{RL} + ( q_{LR}p_R + q_{RL} p_L ) )\delta t + o (\delta t) \\
&=-  (- q_{RL} + ( q_{LR}p_R + q_{RL} p_L )) \delta t +  (- q_{RL} + ( q_{LR}p_R + q_{RL} p_L ) ) \delta t + o (\delta t) \\
&= o(\delta t).
\end{split}
\end{equation}
Letting $q_S = q_{LR}p_R + q_{RL} p_L = (p_{R,S} + p_{L,S}) / \delta t = (p'_{R,S} + p'_{L,S}) / \delta t$, we obtain Eq.~(12) of the main text.

\subsection{Time-scale analysis of the information motor with tape}

For a time $\tau_1$, each cell of the tape couples to the ratchet.
Since $\tau_1$ is short compared to diffusion ($r\tau_1\ll1$), diffusive jumps are very rare.
They occur with a negligible probability $r\tau_1\ll1$ during the interval $\tau_1$.
Thus, during $\tau_1$ all the dynamics are captured by Eq.~(13) in the text.

To obtain the solution of Eq.~(13) for slow adiabatic driving of $F_S(t)$ during the interval $\tau$ with $\gamma\tau_1\gg 1$, we note that in this limit the slow variation of $F_S$ allows the system to relax to the equilibrium associated to $F_S(t)$ at each time $t$.
Thus, the solution at time $t$ in this limit is obtained by setting ${\dot p}_{R,N}(t)={\dot p}_{L,N}(t)=0$ in Eq.~(13) to find the equilibrium distribution with $F_S(t)$:
\begin{equation}
\begin{split}
p_{L,S}(t)&=p_R\frac{1}{1+e^{F_S(t)-\Delta\mu}}\\
p_{R,S}(t)&=p_L\frac{1}{1+e^{F_S(t)}},
\end{split}
\end{equation}
with $p_{R,N}(t)=p_R-p_{L,S}(t)$ and $p_{L,N}(t)=p_L-p_{R,S}(t)$, and having noted $p_N=1$ and $p_S=0$, initially.
Thus the system evolves quasistatically during $\tau_1$ despite the entire process being completed so fast that no diffusive jumps occur.
Finally, at time $\tau_1$, $F_S=f=-\ln(q_{RL}\delta t)$, and using $\Delta\mu=\ln(q_{LR}/q_{RL})$, we find
\begin{equation}
\begin{split}
p_{L,S}&=q_{LR}\delta t p_{R,N} \\
p_{R,S}&=q_{LR}\delta t p_{L,N}.
\end{split}
\end{equation}
Imposing probability conservation, $p_{L,S}(t)=p_Rp_N-p_{R,N}(t)$ and $p_{R,S}(t)=p_Lp_N-p_{L,N}(t)$, and keeping terms only to first order in $\delta t$ we arrive at Eq.~(14) of the main text
\begin{equation}\label{eq:ppost}
\left(\begin{array}{c}
p'_{R,N} \\
p'_{R,S}  \\
p'_{L,N} \\
p'_{L,S}
\end{array}
\right)
=
\left(\begin{array}{c}
(1-q_{LR}\delta t) p_R\\
 q_{RL}p_L\delta t  \\
(1-q_{RL}\delta t) p_L\\
 q_{LR} p_R\delta t
\end{array}
\right).
\end{equation}
We can compare this behavior with the information motor after measurement and feedback.
Equation~(\ref{eq:meas}) above gives the probability density $p_{z,m}$ after measurement, but before the potential is flipped.
Flipping the potential results in an interchange of the probabilities to be in states $R,S$ and $L,S$.
Thus after the switch $p^\prime_{R,S}=p_{L,S}=q_{RL}p_L\delta t $ and $p^\prime_{L,S}=p_{R,S}= q_{LR} p_R\delta t$, in agreement with correlations induced by the tape given in Eq.~(\ref{eq:ppost}) above.
\end{widetext}

\end{document}